# Surveying Instructors' Attitudes and Approaches to Teaching Quantum Mechanics

Shabnam Siddiqui and Chandralekha Singh

*Department of Physics and Astronomy, University of Pittsburgh, Pittsburgh, PA, 15260, USA*

**Abstract**. Understanding instructors' attitudes and approaches to teaching quantum mechanics can be helpful in developing research-based learning tools. Here we discuss the findings from a survey in which 13 instructors reflected on issues related to quantum mechanics teaching. Topics included opinions about the goals of a quantum mechanics course, general challenges in teaching the subject, students' preparation for the course, comparison between their own learning of quantum mechanics vs. how they teach it and the extent to which contemporary topics are incorporated into the syllabus.

## I. INTRODUCTION

While several studies have been conducted on student learning of quantum mechanics (QM) [1-5], an understanding of faculty attitudes and approaches to teaching QM has remained largely anecdotal. A systematic survey of faculty perspectives can be helpful in making further progress in improving QM teaching. Textbooks (and faculty members) follow a variety of approaches: some begin with wave functions, e.g., in the context of the one-dimensional infinite square well, some start with spin and yet others begin with the postulates of quantum mechanics and Dirac notation. These differences may be due to differing opinions among instructors regarding the goals of the course, course content, and the appropriate pedagogical approaches.

Here, we discuss selected findings of a survey about QM teaching given to 13 faculty members, six from the University of Pittsburgh (Pitt) and seven from other institutions. The 21 item survey covered a variety of topics, including the goals of the course, students' mathematical preparation, course content and teaching styles/methods. We also discussed the responses individually with the Pitt instructors. The responses indicate that there is more agreement among instructors on many issues than was expected anecdotally. Still, important variations remained on many of the issues covered in the survey.

## II. GOALS OF A QM COURSE

One of the questions on the survey was the following:

**Feynman once said that nobody understands quantum mechanics. If understanding of QM is so difficult for experts who have spent years learning it, it will be challenging for students who spend a mere semester or two trying to learn it. In your opinion, what are the objectives/goals of an upper level (undergraduate) QM class?**

Interestingly, all thirteen faculty shared common opinion on this issue. According to them, the primary goals of an upper division undergraduate QM course are to learn the postulates and formalism, and gain expertise in applying the formalism to solve diverse problems. For example, one of them noted: *"I think there are a number of goals.
1) ...I think the formalism of QM is "easy". I try to encourage students to realize that this is a great example of a formalism that gives the best predictions for experiments - even if you do not completely understand the formalism.
2) I think a lot of the QM "strangeness" is actually not that strange. Quantization and uncertainty principles are not as hard as I think we make it. The part nobody understands is the measurement issue. I think one goal is to get students to realize where the real fundamental issues are."*

Some noted that the students should understand the relationship between the relevant mathematics and the QM concepts, be able to express the concepts mathematically and gain facility in applying relevant mathematics to relatively simple but important problems. Others also wanted students to get a feel for the numbers, e.g., why explicit quantum effects are not commonly observed in everyday life, why it is so difficult to produce quantum coherence, etc. The following faculty member emphasized the importance of helping students discern the internal consistency of the theory: *".want students to have had, at least at their peak of understanding, a glimpse of the internal*

*consistency of the theory. And in achieving this, they get an important message that there was a process that led to the creation of QM theory. This wasn't the product of some science fiction writer's imagination."*

Faculty member's written responses and individual discussions with some of them suggest that many of them believed that it is not necessary to discuss with students what Feynman was saying. For example, one of them noted: *"I think that the application of quantum mechanics to different situations is easy. I think it can be made easy for students. That is my goal. I want them to understand the rules (which are easy), and to understand how to apply the rules, which is straightforward. I think that Feynman was referring to "where do these crazy rules come from?" I don't get into these philosophical details. I just want the students to understand the rules and know how to apply them. Where the rules 'come from' they can worry about on their own time."* This response is also pragmatic in a similar manner: *"In my opinion the postulates are very beautiful and very complete. I don't expect the students to understand where they come from…I don't understand that either, but I do expect the students to eventually gain an intuitive understanding of what the postulates are, and how to apply them to solve real physical problems"*.

Some also felt that it is unlikely that students can gain a good understanding of QM in a one semester course because of the difficulties involved in learning a subject which is so different from those that students had learned previously. For example, one of the faculty members emphasized: *"Students need many coats of paint to understand QM"*.

## III. MATH PREPARATION

In order to understand the faculty member' views about how mathematically well-prepared students are for taking an upper-level course in QM, we posed the following question: **Rate on a scale of 1-5 (1 means not at all prepared and 5 means extremely well-prepared) how well-prepared students are to take QM in terms of their math preparation in the following mathematical topics (a) Linear algebra, (b) Special functions, (c) Ordinary differential equations, (d) Partial differential equations**.

The data suggest that the mathematical preparation according to most faculty members in each of these categories is either average (rating of 3) or below average (rating of 2 or 1). Regardless of the size of the school, a majority of faculty members noted that students' understanding of special functions is below average (2/3rd of the faculty gave a rating of 2) whereas for topics like linear algebra, partial differential equations and ordinary differential equations, roughly two-thirds of the faculty members gave a rating of 3. Individual discussions with some of the faculty indicate that they felt that improving students' mathematical preparation on these topics may be beneficial in helping students learn QM.

## III. TEACHING STRATEGIES

The objectives for surveying these issues were to learn from the faculty members about their styles, methods and strategies for teaching QM, to identify the possible commonalities between their teaching styles/methods, to learn from them about their views on the difficulties students have in learning QM and how they help students overcome these difficulties.

A. **Instructors' Recollections of Their QM Course**

One of the factors that may influence faculty members in shaping their teaching styles is how they learned when they were students and the strategies and methods used by their instructors in helping them learn. Motivated by how they learned QM, we posed the following question: **Do you remember any difference between your own learning of quantum mechanics vs. other areas of physics? What unique strategies, if any, did you use to learn QM?**

Written responses and individual discussions with some of them suggest that some faculty members struggled to make sense of QM as an undergraduate although they were facile at the mathematical calculations and obtained good grades in the course. For example, one faculty member noted: *"I was completely lost when taking QM as an undergraduate. I could do the math, so I churned out answers while having no conceptual basis for what I was doing. So I guess I really learned QM when I was in grad school, talking to my classmates"*. Another faculty member noted the gap between having a knowledge structure to discern how things fit together and the grades obtained: *"Yes. Everything else that I learned was straightforward. QM made no sense until graduate school. Even though I got A's on every test, I really had no clue how it held together. I teach it completely different from how I learned it"*.

Very few (one) faculty members noted that they enjoyed learning QM. A few (two) explicitly mentioned that they found it difficult to focus on the concepts or that they were bogged down in the mathematics. Some (five) mentioned working very hard and relying on teaching themselves from the textbook because they could not make sense of the instructor. A few faculty members had developed effective strategies to learn QM such as the following: *"For me, it really came in two stages. I focused first*

*on executing the mechanics and then trying to understand what it "meant"".*

### B. Examples of Common Difficulties

We also investigated instructors' views about some of the difficulties their students have in learning QM by asking the following question: **From your experience can you tell us about some of the specific difficulties students have in learning QM?**

Written responses and individual discussions with some of them suggest that despite the fact that the faculty members rated students' mathematical preparation to be inadequate; their major concern often was students' conceptual difficulties in learning QM. One faculty member noted that learning the rules is easy: *"I find that QM more than any other subject has two distinct levels 1) Apply the rules. 2) formulate your own mental picture of what is physically happening, i.e. have an understanding of the concepts. To me, the rules are probably the most straightforward of any physics discipline but the concepts are perhaps the hardest. This large discrepancy causes students difficulty".*

Moreover, some of them felt that the difficulty may be exacerbated by inaccurate descriptions in modern physics courses. For example, one of them noted: *"... there is lots of baggage from modern physics courses and the stories they learn from there"*. In discussion with one faculty member, he noted that in his opinion the misconceptions in introductory classical mechanics are often unavoidable because people try to make sense of their everyday experiences and the laws of physics do not conform to their naïve interpretations of physical phenomena. On the other hand, he felt that the misconceptions in QM in many situations are avoidable if we do not teach students inaccurate descriptions of quantum systems which is especially common in the modern physics courses.

### C. Making Sense of the Wavefunction

In order to gain insight into how they teach a particular topic, we asked the faculty members about how they help students make sense of the wavefunction. The opinions on this issue varied. Five out of thirteen focused on helping students learn to draw the wave function in position space and learn to use it to answer questions about measurements of physical observables and expectation values. For example, one faculty noted: *"For cases where the spatial part of the wavefunction is real, we do a lot of sketching, relating the curvature to E-V. I give them some hypothetical messy potential and they have to sketch a wavefunction that is at least plausibly a solution"*. The other six of the faculty admitted that they had not explicitly thought about this issue as in the following response: *"I haven't thought about this per se. I guess the main thing is focusing on the connection to measurements. I focus much more on the probability of making measurements than on the expectation value"*.

Written responses and individual discussions suggest that two of the faculty stressed in terms of the state of the system being a vector in the Hilbert space. The following is an example along this line: *"I tell them that Hilbert space is a porcupine. Infinite dimensional space is a porcupine with each spine perpendicular to the other. Each spine represents an eigenstate. A wave function is a vector pointing in an arbitrary direction in the porcupine"*.

### D. Classical Mechanics and Quantum Mechanics

By the time students take QM, they already have well-developed classical intuition. But classical mechanics is a deterministic theory whereas QM is probabilistic. Therefore, it is important that students understand the similarities and differences between the two theories. In order to learn from the faculty members about the approaches they use to help the students see the connections and differences between classical mechanics and QM, we posed the following question: "**In your class, what types of connections do you make between quantum physics and classical mechanics (CM)? Also, how do you explain the differences between QM & CM? Please give specific examples**".

Responses suggest that the faculty members have varied opinions on the extent to which the connections and differences between CM and QM should be emphasized. For example, the following response from a faculty member suggests that he does not make any significant effort to discuss the connections: *" I do not emphasize the connections between QM and classical physics. I find this mostly distracting. . I use CM as a "foil" and emphasize the differences that QM has with CM. I think it's more important to understand the differences rather than the similarities (since students already have a good CM intuition, but tend to have no QM intuition, and these intuitions are very very different…)"*. Several other faculty members also mainly focused on the differences as expressed in this comment: *"Perhaps the main emphasis in this regard is a discussion in terms of how the state of a system is characterized. In CM we tend to characterize the state of the system by the position and velocity and how this determines energy, etc. In QM, the wavefunction characterizes the state. From there, the goal is to predict measurements, and we discuss the differences in this process"*. Written responses and individual discussions suggest that those who believed that it is important to make these connections often felt that they will help students develop intuition.

Another question that the faculty members were asked was whether they agree or disagree that the semi-classical models such as the Bohr's model of the hydrogen atom can be misleading for students learning QM. The instructors' level of support for the model in general varied. A2/3 of the faculty felt that the model is misleading and can lead to the students developing misconceptions that will be difficult to eliminate. Here is an example: *"I find the Bohr model hideous. I really wish modern physics classes could go beyond the Bohr model some day, so that we don't have to unteach it later."* Others (1/3) felt that it has a place in a modern physics course as displayed in the following comment: *"I do not teach the Bohr model at all in quantum courses. It has its place in elementary modern physics courses, but quantum students should realize that energies come from diagonalizing the Hamiltonian, and wavefunctions do not really describe classical trajectories."* One faculty member noted that he is not against the semi-classical models but avoids the Bohr's model: *"...I avoid it. At least that's how I feel when teaching QM (although out in the real world I appreciate that semiclassical ideas can be useful, like understanding how electrons move in solids), but I avoid semiclassical stuff when teaching QM."* Another faculty member felt that the Bohr's model is not a semi-classical model: *"I definitely think the only real value of the Bohr model is to show how physics works as an experimental science - it was a great starting point as it was able to explain certain measurements. But, once more measurements are made (especially related to angular momentum, which it gets wrong) the model fails, and you need a different model. I would say the real misleading thing is to call it semi-classical. There are semi-classical techniques that work in some situations... Bohr model is just wrong."*

**E. Introduction of Novel Developments in QM**

In the last few decades, major experimental and theoretical advances have been made that elucidate foundational issues in quantum mechanics. Therefore, faculty members were asked to suggest any novel developments in the field of QM that offered new insights in resolving some of the foundational issues and whether they would discuss them with students. Quantum information/ computation, entanglement, EPR paradox, Bell inequality and its experimental confirmation topped the list. Here are responses from three instructors:

*"Quantum information theory should be introduced more to QM. Ideas of qubits and manipulating them should be made more important. Ideas of partial measurement, projections, could be useful."*

*"Being able to do single-photon experiments with students is new - although the experiments themselves aren't new. It seems like quantum optics is a place where some new experiments are revealing that QM continues to predict very strange things correctly. I am completely open to discussing new things with students but there's not much room in the course..."*

*"The Bell inequality experiments are certainly a major breakthrough that should be covered in a quantum course if there is enough time to do it properly (which unfortunately is often not the case)."*

## IV. SUMMARY

It appears that all of the faculty members share common opinions about the goals of an upper-division undergraduate QM course, although there were some differences of opinion on various issues. Another interesting finding is that regardless of the type of an institution, instructors on average considered students' mathematical preparation for QM to be below average. Moreover, if the faculty members' reflections about their QM courses suggest that many of them were having difficulty in making sense of their calculations, it is important to consider strategies to bridge the gap between the conceptual and quantitative aspects of QM especially because the diversity of students' prior preparation in the classrooms has increased greatly over the years. We also observed that for a particular question, the responses of the faculty from the University of Pittsburgh were similar, whereas responses of faculties from other institutions were diverse. . In conclusion, we can say that surveying a small set of faculty gave us clues towards the real issues in teaching upper level quantum mechanics. Also, this survey suggests that attitudes of faculty are built on the needs of their respective institutes.

## ACKNOWLEDGMENTS
We thank participating faculty and NSF for support.